\documentclass{emulateapj}
\usepackage{apjfonts}

\newcommand{\Msun}      {\mbox{$\rm\,M_{\mathord\odot}$}}

\begin{document}

\def\lsim{\mathrel{\lower .85ex\hbox{\rlap{$\sim$}\raise
.95ex\hbox{$<$} }}}
\def\gsim{\mathrel{\lower .80ex\hbox{\rlap{$\sim$}\raise
.90ex\hbox{$>$} }}}

\lefthead{XTE J2123--058 in Quiescence}
\righthead{J.A. Tomsick et al.}

\title{The Low Quiescent X-Ray Luminosity of the Neutron Star Transient 
XTE J2123--058}

\author{John A. Tomsick\altaffilmark{1},
Dawn M. Gelino\altaffilmark{1}, 
Jules P. Halpern\altaffilmark{2}, 
Philip Kaaret\altaffilmark{3}}

\altaffiltext{1}{Center for Astrophysics and Space Sciences, Code
0424, University of California at San Diego, La Jolla, CA,
92093, USA (e-mail: jtomsick@ucsd.edu)}

\altaffiltext{2}{Columbia Astrophysics Laboratory, Columbia University,
550 West 120th Street, New York, NY 10027}

\altaffiltext{3}{Harvard-Smithsonian Center for Astrophysics, 60 Garden Street,
 Cambridge, MA, 02138, USA}

\begin{abstract}

We report on the first X-ray observations of the neutron star soft 
X-ray transient (SXT) XTE J2123--058 in quiescence, made by the 
{\em Chandra X-ray Observatory} and {\em BeppoSAX\/}, as well as 
contemporaneous optical observations.  In 2002, the {\em Chandra} 
spectrum of XTE J2123--058 is consistent with a power-law model, 
or the combination of a blackbody plus a power-law, but it is not 
well-described by a pure blackbody.  Using the interstellar value 
of $N_{\rm H}$, the power-law fit gives $\Gamma = 3.1^{+0.7}_{-0.6}$ 
and indicates a 0.3--8 keV unabsorbed luminosity of 
$(9^{+4}_{-3})\times 10^{31}$ ($d$/8.5 kpc)$^{2}$ ergs~s$^{-1}$ 
(90\% confidence errors).  Fits with models consisting of thermal plus 
power-law components indicate that the upper limit on the temperature 
of a 1.4\Msun, 10~km radius neutron star with a hydrogen atmosphere is 
$kT_{\rm eff} < 66$~eV, and the upper limit on the unabsorbed, bolometric 
luminosity is $L_{\infty} < 1.4\times 10^{32}$ ergs~s$^{-1}$, assuming
$d = 8.5$~kpc.  Of the neutron star SXTs that exhibit short ($<$1 year) 
outbursts, including Aql X-1, 4U 1608--522, Cen X-4, and SAX J1810.8--2609, 
the lowest temperatures and luminosities are found for XTE J2123--058 and 
SAX J1810.8--2609.  From the {\em BeppoSAX\/} observation of 
XTE J2123--058 in 2000, we obtained an upper limit on the 1--10 keV 
unabsorbed luminosity of $9\times 10^{32}$ ergs~s$^{-1}$.  Although 
this upper limit allows that the X-ray luminosity may have decreased 
between 2000 and 2002, that possibility is not supported by our 
contemporaneous $R$-band observations, which indicate that the optical 
flux increased significantly.  Motivated by the theory of deep crustal 
heating by Brown and co-workers, we characterize the outburst histories 
of the five SXTs.  The low quiescent luminosity for XTE J2123--058 is 
consistent with the theory of deep crustal heating without requiring 
enhanced neutron star cooling if the outburst recurrence time is 
$\gsim$70 years.  

\end{abstract}


\section{Introduction}

Accreting neutron stars can be found in high-mass (HMXB) or low-mass (LMXB) 
X-ray binary systems.  The majority of HMXBs have transient X-ray emission.  
Their outburst spectra are relatively hard and X-ray pulsations from these 
highly magnetized ($B\sim 10^{12}$~G) neutron stars are typically detected.  
A wide variety of X-ray behaviors are seen for neutron star LMXBs, but, in 
general, the lack of X-ray pulsations from most (but not all) of these systems, 
and the emission of type I X-ray bursts from some, suggest that they harbor 
neutron stars with relatively low magnetic field strengths ($B\sim 10^{8-9}$~G).  
During $\approx 33$ years of X-ray observations, some sources (e.g., Sco X-1, 
Cyg X-2) have maintained luminosities approaching the Eddington limit of 
$\approx 10^{38}$ ergs~s$^{-1}$, while others are able to maintain persistent 
luminosities several orders of magnitude lower \citep{wilson03}.  In addition, 
there is a class of transient neutron star LMXBs for which the luminosity 
varies from a substantial fraction of Eddington to quiescent levels typically 
near $10^{32-33}$ ergs~s$^{-1}$.  In outburst, these systems have relatively 
soft spectra compared to the HMXBs, and are commonly grouped with black hole 
transients as soft X-ray transients (SXTs).  

In quiescence, most neutron star SXTs exhibit X-ray energy spectra with a 
component that is typically fitted well by a blackbody, suggesting that 
the origin of this component is thermal emission from the surface of a 
cooling neutron star.  Although a pure blackbody often provides a good fit 
to the spectrum, unphysical neutron star radii near 1 km are inferred unless 
an atmosphere is modeled \citep{rutledge99}.  In addition to the thermal 
component, the energy spectra often contain a second component that has a 
power-law shape.  The brightest and best studied systems in this class, 
Cen X-4 and Aql X-1, usually display both components 
\citep{rutledge01,rutledge02a,cs03,campana04}.  However, other systems may 
be dominated by the thermal component, such as MXB 1659--29 \citep{wijnands03b} 
and sources X-5 and X-7 in the globular cluster 47 Tucanae \citep{heinke03}, 
or by the power-law component, such as SAX J1808.4--3658 \citep{campana02} 
and EXO 1745--248 \citep{wijnands03a}.  Although theories for the thermal 
component, such as the deep crustal heating model of \cite{bbr98}, are 
relatively well-developed and are being tested with observations, the origin 
of the power-law component is not understood beyond suggestions that it may 
be related to accretion onto the neutron star magnetosphere \citep{campana98} 
or a putative pulsar wind colliding with infalling matter from the companion 
star \citep{tavani91}.  In addition to our lack of understanding of the 
power-law component, questions remain about the mass accretion rate in 
quiescence, the origin of rapid (100--10,000~s) variability 
\citep{rutledge02a,campana04}, and the origin of variability in the thermal 
component on longer time scales \citep{rutledge02a}.  Another important 
question is if quiescent observational properties correlate with other 
known differences between neutron star SXTs, such as whether they are 
millisecond X-ray pulsars (during outbursts) or not, whether the systems 
are in the field or in globular clusters, and whether their X-ray 
outbursts are long (years to decades) or short (weeks to months).

Here, we report on X-ray and optical observations of the field neutron 
star SXT XTE J2123--058 taken during quiescence.  XTE J2123--058 had its 
only detected X-ray outburst in 1998 June-August \citep{lss98,tomsick99}, 
and we focus on observations made with {\em Chandra}, {\em BeppoSAX\/}, 
and optical telescopes 2--4 years after the outburst.  During the outburst, 
the {\em Rossi X-ray Timing Explorer (RXTE\/)} detected type I X-ray bursts 
and a pair of kHz quasi-periodic oscillations \citep{homan99,tomsick99}, 
indicating that the system contains a rapidly rotating neutron star.  
However, coherent X-ray pulsations were not found.  The 6~hr binary 
orbital period and the fact that the binary inclination of the system 
is relatively high were established from optical modulation and the 
presence of partial eclipses in the optical light curve 
\citep{tomsick99,swg99,zurita00,shahbaz03}.  The high Galactic latitude 
($b = -36^{\circ}$) and low extinction have allowed for detailed optical 
studies of XTE J2123--058 in quiescence even though the source is rather 
faint at its relatively large distance ($8.5\pm 2.5$ kpc).  The
optical observations show that XTE J2123--058 consists of a K7~V star on 
or close to the main sequence and a neutron star for which mass 
determinations of $1.5\pm 0.3$~\Msun~\citep{tomsick01,casares02,tomsick02}
and 1.04--1.56\Msun~\citep{shahbaz03} have been obtained.  The focus
of this paper is the first X-ray study of XTE J2123--058 in quiescence.

\section{Observations and Analysis}

We observed XTE J2123--058 with {\em Chandra} on UT 2002 November 13
(ObsID 2709), using the Advanced CCD Imaging Spectrometer (ACIS) with 
the target placed on one of the back-illuminated ACIS chips (ACIS-S3).  
For our analysis, we used the ``level 2'' event list produced by the 
standard data processing with ASCDS version 6.9.2 using Calibration Data 
Base (CALDB) version 2.17.  Light curves using counts from the full 
field-of-view do not show any background flares, allowing us to use the 
data from the full 17,706~s integration.  Using the {\em Chandra} 
Interactive Analysis of Observations (CIAO) version 3.0 software routine 
{\em wavdetect} \citep{freeman02}, we searched for sources on the S3 chip 
in the 0.3--8 keV energy band.  We detected 22 sources with counts between 
5 and 108 per source, using a detection threshold of $10^{-6}$, including 
a 24 count source at R.A. = $21^{\rm h}23^{\rm m}14^{\rm s}\!.54$, 
decl. = $-05^{\circ}47^{\prime}53^{\prime\prime}\!.2$ (equinox J2000, 
uncertainty $0^{\prime\prime}\!.6$).  This position is consistent with the 
target's optical position \citep{tomsick99}, and we conclude that this source 
is the quiescent X-ray counterpart of XTE J2123--058.

We also report on a {\em BeppoSAX\/} observation of the XTE J2123--058 field 
made on UT 2000 May 12--13.  We produced 1--10 keV images using the data from 
the two Medium Energy Concentrator/Spectrometer (MECS) units that were 
operational during the observation (units 2 and 3).  We also produced a 
0.1--10 keV image using data from the Low Energy Concentrator/Spectrometer 
(LECS).  We obtained a MECS exposure time of 46,340~s and a LECS exposure
time of 17,080~s.  To search for sources, we convolved each of the three 
images with a two-dimensional Gaussian with a width ($\sigma$) of 5 pixels 
($40^{\prime\prime}$) in both directions.  Only one source was clearly 
detected in both of the MECS images at R.A. = $21^{\rm h}22^{\rm m}50^{\rm s}\!.6$, 
decl. = $-05^{\circ}45^{\prime}09^{\prime\prime}$ (equinox J2000, 
uncertainty $\sim 1^{\prime}$), and this source, which we call 
SAX J2122.8--0575, is also present in the LECS image.  It is clear that 
SAX J2122.8--0575 is not XTE J2123--058 as they are separated by
$6^{\prime}\!.6$.  We conclude that XTE J2123--058 was not detected during 
the {\em BeppoSAX\/} observation, and we derive an upper limit on its X-ray 
flux below.

Finally, we obtained $R$-band images on three occasions close to the times of 
the X-ray observations.  As shown in Table~\ref{tab:obs}, we observed 
XTE J2123--058 using the 2.4~m Hiltner telescope of the MDM Observatory on 
2000 July 24, about 2 months after the {\em BeppoSAX\/} observation.  We also 
observed XTE J2123--058 with the Shane 3~m telescope of Lick Observatory about 
2 months before the {\em Chandra} observation and again at MDM about 2 weeks 
after the {\em Chandra} observation.  For both MDM runs, we used the same 
SITe $2048\times 2048$ pixel, thinned, back-illuminated CCD with a spatial 
scale of $0^{\prime\prime}\!.275$ per 24~$\mu$m pixel, and an $R$ filter that 
is very close to Harris $R$. 
\footnote{See http://www.astro.lsa.umich.edu/obs/mdm/technical/filters
for the exact transmission curve.}  At Lick, we used the Prime Focus Camera,
with a SITe $2048\times 2048$ pixel, thinned CCD with a spatial scale of 
$0^{\prime\prime}\!.296$ per 24~$\mu$m pixel, and a Kron-Cousins $R$ filter.  
In all, we obtained thirteen 600--700~s exposures (see Table~\ref{tab:obs}), 
and we reduced the images using standard IRAF \footnote{IRAF (Image Reduction 
and Analysis Facility) is distributed by the National Optical Astronomy 
Observatories, which are operated by the Association of Universities for 
Research in Astronomy, Inc., under cooperative agreement with the National 
Science Foundation.} routines.  For XTE J2123--058, we carried out the 
photometry with the IRAF package {\em phot} and used two calibrated reference 
stars with $R$ magnitudes of 19.47 and 19.51.  In 2002, the conditions at 
Lick were photometric, and we observed \cite{landolt92} standards to obtain 
the calibration.  We note that this calibration is about 0.1 magnitudes 
brighter than the calibration previously obtained using MDM observations 
from 1998 reported in \cite{tomsick99}.

\section{Energy Spectrum and Source Luminosity}

We used the CIAO software routine {\em psextract} to produce the
ACIS energy spectrum for XTE J2123--058 and to create the appropriate
instrument response matrix for the spectrum.  The software used
CALDB 2.23 to create the response matrix, and we included a correction 
for the time-dependence of the ACIS response.  The source spectrum 
included counts from a circular region with a 5 pixel (about 
$2^{\prime\prime}\!.5$) radius, and we estimated the background level 
using counts from a source-free annulus around the target position.  
The source spectrum consists of 24 counts, and we estimate a background 
level of 0.7 counts in the extraction region.  We produced a ``light curve''
of the source in six time bins of $\approx 3000$~s.  Each bin contains 
between 2 and 6 counts, which is consistent with a constant flux; however, 
the low count rate does not allow us to place tight constraints on the 
possible amplitude of variability.  The small number of counts also 
indicates that $\chi^{2}$ statistics are not appropriate for spectral 
analysis, as the assumption of a Gaussian probability distribution in 
each spectral bin is not met.  Thus, we carried out our spectral analysis 
by minimizing the Cash statistic \citep{cash79}, which is appropriate in 
cases where the assumption of a Poisson probability distribution in each 
spectral bin is valid.  We fitted the spectrum using XSPEC 11.2.

The spectra of other neutron star SXTs are typically well-described
by a blackbody, a power-law, or both components with interstellar 
absorption.  We began by fitting the spectrum for XTE J2123--058 with an 
absorbed power-law, and the results are given in Table~\ref{tab:spectra}.  
Once we found the best-fitted parameters by minimizing the Cash statistic, 
we determined the quality of the fit and the 90\% confidence errors on the 
parameters by producing and fitting 10,000 simulated spectra.  We used 
the best-fitted parameters from the fits to the actual data as input to the 
simulations.  Using simulations to determine the parameter errors is 
necessary only because of the small number of counts in the spectrum.  We 
tested our method by producing spectra with 2--3 times as many counts, 
and we found that the errors produced by calculating changes in the 
Cash statistic (the standard technique) match the values we obtain 
using simulations.  In addition to the power-law model, we used simulations 
to determine the errors on parameter for all other fits presented in this work.  
For the power-law model, we obtain $N_{\rm H} = (7.0^{+40.0}_{-7.0})\times 
10^{20}$ cm$^{-2}$ for the column density and $\Gamma = 3.1^{+2.8}_{-0.8}$ 
for the photon index.  The fit using a blackbody model is significantly
worse than the power-law model as indicated by the fact that for 97\% of the
simulated spectra we obtained a better Cash statistic than the one
obtained when fitting the actual spectrum (compared to 69\% for
the power-law model).

\begin{figure}
%\plotone{f1.ps}
\centerline{\includegraphics[width=0.52\textwidth]{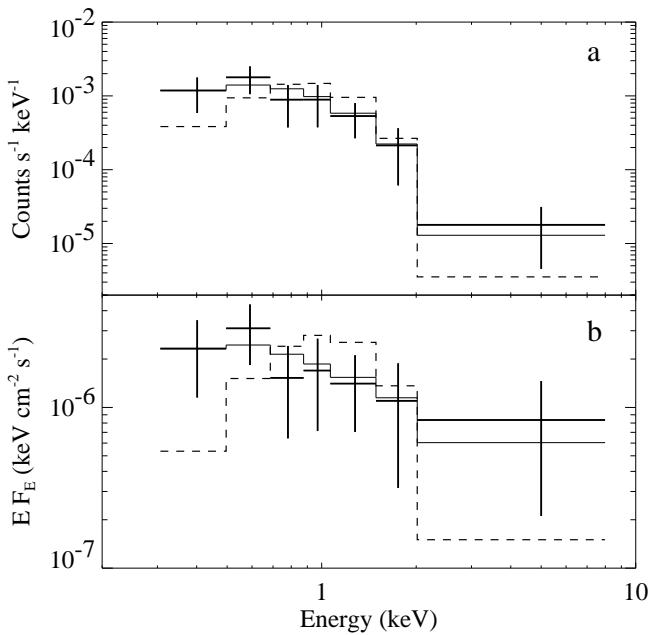}}
%\vspace{0.2cm}
\caption{{\em Chandra}/ACIS energy spectrum of XTE J2123--058. {\it Top}: folded
through the detector response. {\it Bottom}: unfolded.  In each panel, the 
{\it solid line} is a power-law fit to the data (using Cash statistics as described 
in the text) with $N_{\rm H}$ fixed at the interstellar value.  The power-law 
photon index is $3.1^{+0.7}_{-0.6}$, and the unabsorbed 0.3--8 keV luminosity is 
$(9^{+4}_{-3})\times 10^{31}$ ergs~s$^{-1}$.  A blackbody model ({\it dashed line}), 
by itself, does not provide a good description of the spectrum.\label{fig:spectrum}}
\end{figure}

We refitted the spectrum with the same spectral models but with
$N_{\rm H}$ fixed to the interstellar column density.  The interstellar 
$N_{\rm H}$ comes from the $A_{V}$ measurement of \cite{hynes01}, which gives 
$N_{\rm H} = (6.6\pm 2.7)\times 10^{20}$ cm$^{-2}$, and the total Galactic
\ion{H}{1} value of $5.7\times 10^{20}$ cm$^{-2}$ from \cite{dl90}.
We adopt a value of $6\times 10^{20}$ cm$^{-2}$.  For the power-law alone, we 
obtain $\Gamma = 3.1^{+0.7}_{-0.6}$ and an unabsorbed 0.3--8~keV flux of 
$(1.1^{+0.5}_{-0.4})\times 10^{-14}$ ergs~cm$^{-2}$~s$^{-1}$, which corresponds 
to a luminosity of $(9^{+4}_{-3})\times 10^{31}$ ergs~s$^{-1}$ at a distance of 
8.5 kpc.  The best fitted blackbody temperature is $240^{+60}_{-50}$ eV, but the 
quality of the blackbody fit is even worse (98.7\%) with $N_{\rm H}$ fixed.  
Figure~\ref{fig:spectrum} shows the spectrum (rebinned for clarity, although 
the Cash fits are performed without rebinning) along with the best fitted power-law 
and blackbody models with $N_{\rm H}$ fixed.  It is clear that the curvature 
of the blackbody model is too large, underpredicting the observed spectrum 
below 0.7 keV and above 2 keV.  This, along with the measurements of the 
fit quality from the simulations, indicates that the spectrum is well-described 
by a power-law model or the combination of a blackbody and a power-law model, 
but it is not well-described by a blackbody alone.

Although a power-law alone provides an adequate description of the ACIS
spectrum, we performed additional fits to obtain limits on the 
temperature and luminosity of the putative thermal component that 
is expected to be emitted from the surface of the neutron star.  We fitted
the spectrum with a model consisting of a power-law component along with
thermal emission from an atmosphere composed of hydrogen.
The latter component was modeled using the ``Neutron Star Atmosphere''
(NSA) model of \cite{zps96}.  We assumed a neutron star mass of 
1.4\Msun, which is consistent with the measured mass, and a radius of 
10~km.  With these parameters fixed, the remaining free parameters in 
the NSA model are the temperature ($kT_{\rm eff}$) and the distance.  
For XTE J2123--058, there are several arguments that lead to distance 
estimates in the range 5--15 kpc \citep{tomsick99,homan99,zurita00}, 
but the most reliable estimates come from optical observations of the 
source in quiescence.  Previous estimates include $8\pm 3$ kpc 
\citep{zurita00}, $8.5\pm 2.5$ kpc \citep{tomsick01}, and $9.6\pm 1.3$ kpc 
\citep{casares02}.  Here, we adopt a range of 6--11~kpc.  We performed 
the NSA plus power-law fits with the distance fixed to three values 
spanning this range (6, 8.5, and 11 kpc).  As shown in Table~\ref{tab:nsa}, 
the 90\% confidence upper limits on $kT_{\rm eff}$ are 57, 66, and 73 eV, 
respectively, for these three distances.  Using $L_{\infty} = 
4\pi R^{2}\sigma T_{\rm eff}^{4} (1-2GM/Rc^{2})$, where $R$ = 10 km and
$M$ = 1.4\Msun, upper limits on the unabsorbed luminosity from the NSA 
component as seen by a distant observer are $8.0\times 10^{31}$, 
$1.4\times 10^{32}$ and $2.1\times 10^{32}$ ergs~s$^{-1}$ for distances 
of 6, 8.5, and 11 kpc, respectively.  While these values are bolometric 
luminosities, it should be noted that we cannot rule out the possibility 
that there is thermal emission at energies below the {\em Chandra} bandpass.

Although XTE J2123--058 was not detected by {\em BeppoSAX\/} in 2000 May, 
we can calculate upper limits on its luminosity if we assume that the 
energy spectrum was similar to that measured by {\em Chandra}.  To 
calculate the upper limits, we assume the simplest model of an absorbed 
power-law with a photon index of 3.1 and the interstellar column density.  
For the MECS (units 2 and 3 combined), we measure a 1--10 keV count rate 
of 1.684 ks$^{-1}$ in a circle of radius $2^{\prime}$ centered on the 
XTE J2123--058 position.  This is not significantly higher than the 
expected background rate (using a blank-sky pointing) of 1.675 ks$^{-1}$.
These values imply a 3-$\sigma$ upper limit on the count rate from
XTE J2123--058 of 0.676 ks$^{-1}$, which corresponds to an unabsorbed 
flux $< 1.02\times 10^{-13}$ ergs~cm$^{-2}$~s$^{-1}$, and a 1--10~keV 
luminosity $< 9\times 10^{32}$ ergs~s$^{-1}$ for $d=8.5$~kpc.  For the 
LECS, the 0.1--1 keV count rate in a $2^{\prime}$ radius circle centered 
on XTE J2123--058 is 1.230 ks$^{-1}$, which is actually somewhat lower 
than the expected background rate of 1.466 ks$^{-1}$.  Using the expected
background rate, the 3-$\sigma$ upper limit on the unabsorbed 0.1--1 keV 
flux is $1.75\times 10^{-12}$ ergs~cm$^{-2}$~s$^{-1}$, corresponding to 
a luminosity $< 1.5\times 10^{34}$ ergs~s$^{-1}$ for $d=8.5$~kpc.

\section{Optical Results}

\begin{figure}
%\plotone{f2.ps}
\centerline{\includegraphics[width=0.58\textwidth]{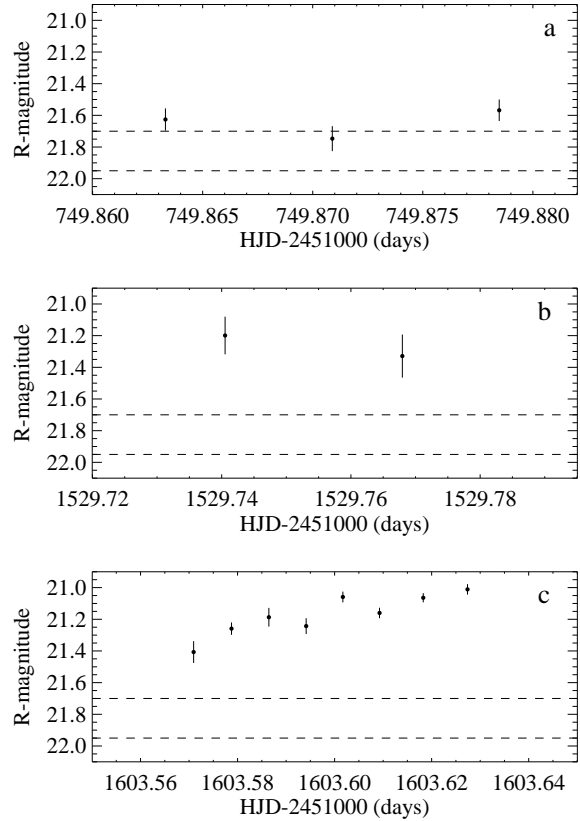}}
\vspace{-2.0cm}
\caption{$R$-band measurements of XTE J2123--058. ({\it a}) 2000 July at MDM 
Observatory. ({\it b}) 2002 September at Lick Observatory. ({\it c}) 2002 November 
at MDM Observatory.  The {\it dashed lines} delimit the levels measured by 
\cite{shahbaz03} during 1999--2000.  An increase in the $R$-band flux between 2000 
and 2002 is apparent.\label{fig:lc_optical}}
\end{figure}

Figure~\ref{fig:lc_optical} shows the $R$ magnitudes for XTE J2123--058 
from three exposures taken in 2000 July at MDM, two exposures taken in 
2002 September at Lick, and eight exposures taken in 2002 November at 
MDM.  Previously, \cite{shahbaz03} reported on extensive $R$-band photometry
of XTE J2123--058 taken between 1999 June and 2000 August.  The $R$-band 
light curve was relatively stable at that time, showing orbital ellipsoidal 
modulations with a peak-to-peak amplitude of about 0.25 magnitudes.  The 
dashed lines in Figure~\ref{fig:lc_optical} indicate the range of the 
modulation in 1999--2000, which was $21.70 < R < 21.95$.  While our 2000 
May measurements of $21.63\pm 0.07$, $21.75\pm 0.08$, and $21.57\pm 0.07$ 
are consistent with the brighter end of this range, our later observations 
indicate that XTE J2123--058 brightened between 2000 and 2002.  The range 
of $R$ between $21.41\pm 0.07$ and $21.01\pm 0.03$ over 25\% of the 6~hr 
binary orbital period during the 2002 November run indicates that the 
source continues to vary, but with a peak level that is at least 0.6--0.7 
magnitudes brighter than the peak level of $R=21.7$ measured by 
\cite{shahbaz03}.  As our observations do not cover the full binary orbit, 
it is not clear whether this change is due to the addition of a constant 
component to the light curve or if the shape of the modulations has changed.  
Although it would be useful to know the orbital phases corresponding to the 
2002 data, the binary ephemeris is too uncertain to extrapolate to 2002 as 
discussed in \cite{tomsick02}.  In the future, it may be worthwhile to 
obtain an X-ray observation while measuring the optical or IR light curve 
for a full 6~hr binary orbit.

As we are reporting the first sensitive X-ray observations of XTE J2123--058 
in quiescence, comparison can be made with a prior prediction for the 
quiescent X-ray luminosity of this source that was made by modeling quiescent 
$R$-band light curves in 1999 and 2000.  \cite{shahbaz03} predicted an X-ray 
luminosity of $\sim$$10^{33}$ ergs~s$^{-1}$, which is only slightly higher 
than our upper limit from the {\em BeppoSAX\/} observation made in 2000, but 
an order of magnitude higher than we observed with {\em Chandra} in 2002.  
While the X-ray observations would allow for the possibility that the 
quiescent X-ray luminosity decreased significantly between 2000 and 2002, 
our optical observations made in 2000 and 2002 indicate that the $R$-band 
flux actually increased over this time.  If the 2000 $R$-band light curve 
showed a significant contribution from X-ray heating, it is difficult to see 
how a drop in X-ray flux could lead to an increase in the optical, thus, it 
is possible that \cite{shahbaz03} over-estimated the contribution from X-ray 
heating.  

\section{Interpretation}

X-ray observations of neutron star SXTs in quiescence provide tests of 
theoretical models for the thermal component.  According to the theory of 
deep crustal heating by \cite{bbr98}, the temperature of the neutron star 
core is maintained by nuclear reactions in the deep crust that occur when 
the mass accretion rate is high during outburst.  As the thermal time scale 
for the core is $\sim$10,000~years \citep{colpi01}, the level of quiescent 
thermal emission is set by the average mass accretion rate over this time 
span according to $L_{\rm q} = 9\times 10^{32}$ $<$$\dot{M}$$>_{-11}$~ergs~s$^{-1}$
(see Equation 1 of Rutledge et al. 2002b\nocite{rutledge02b}), assuming 
1.45~MeV of heat deposited in the crust per accreted nucleon \citep{hz90}.
Here $<$$\dot{M}$$>_{-11}$ is the mass accretion rate averaged over the 
thermal time scale of the core in units of $10^{-11}$\Msun~yr$^{-1}$,
and $L_{\rm q}$ is the quiescent bolometric luminosity.  While the 
time-averaged mass accretion rate is only a predictor of $L_{\rm q}$ if 
the neutron star core has reached thermal equilibrium, the lifetimes
of LMXB systems are much longer than the thermal time scale of the 
core, and it is expected that thermal equilibrium has been established
for all or nearly all of the LMXBs.  While the long thermal time scale
for the core precludes appreciable changes in the core temperature during 
a single SXT outburst, the neutron star crust can be heated significantly 
for the systems with longer outbursts (years to decades) so that the 
quiescent emission is determined by the evolution of the physical conditions 
in the crust \citep{rutledge02b}.  However, for the systems with shorter 
outbursts (weeks to months), the properties of the quiescent thermal 
emission are primarily set by the conditions in the neutron star core 
\citep{bbr98}.  

For a system like XTE J2123--058 that has undergone one $\approx 40$ day 
outburst, the deep crustal heating theory implies that quiescent thermal 
emission provides information about the neutron star core.  Here, we compare 
the quiescent properties of XTE J2123--058 to similar systems.  The 
comparison group includes field neutron star SXTs with outbursts lasting 
less than 1~year for which quiescent X-ray observations have been reported.
We only consider LMXB systems from which X-ray bursts have been detected, 
proving that the accreting object is a neutron star.  We restrict the comparison
group to field systems because the outburst histories for most transients in
globular clusters are uncertain due to source confusion in instruments with 
angular resolution worse than {\em Chandra}.  Thus, our comparison
group consists of the neutron star SXTs Aql X-1, 4U 1608--522, Cen X-4, and
SAX J1810.8--2609.  We also compare the quiescent properties of XTE J2123--058 
to those of the millisecond X-ray pulsar SAX J1808.8--3658 as the quiescent
X-ray luminosities of the two sources are comparable.

Table~\ref{tab:comparison} compares the parameters of the thermal component 
for the five systems, including $kT_{\rm eff}$, $R$, $L_{\infty}$, and the 
gravitational redshift parameter, $g = \sqrt{1-2GM/Rc^{2}}$, assuming that 
all the neutron star masses are 1.4\Msun.  For the spectral fits to 
XTE J2123--058, we set $R = 10$~km, and \cite{jwv04} make the same assumption
for SAX J1810.8--2609.  However, for the other sources, we use the best
fitted radii, and the parameters are taken from 
\cite{rutledge99,rutledge01,rutledge02a}.  The thermal component is 
detected for Cen X-4, Aql X-1, and 4U 1608--522, and a power-law component 
is also required for Cen X-4 and for some of the observations of Aql X-1.
For XTE J2123--058 and SAX J1810.8--2609, the thermal component is not
statistically required, and we report upper limits on the thermal component 
parameters for these two systems.  Also, for all the sources we assume the 
best current distance measurement.  The temperature upper limits for 
XTE J2123--058 (66~eV) and SAX J1810.8--2609 (72~eV) are about a factor of 
2.5 lower than the 4U 1608--522 measurement of 170~eV, and the luminosity 
upper limits for XTE J2123--058 and SAX J1810.8--2609 are about a factor 
of 50 lower than the maximum quiescent luminosity measured for Aql X-1.  
The temperature and luminosity of Cen X-4 are intermediate between the 
hottest and most luminous sources (Aql X-1 and 4U 1608--522) and the coolest 
and least luminous sources (XTE J2123--058 and SAX J1810.8--2609).

\begin{figure}
%\plotone{f3.ps}
\centerline{\includegraphics[width=0.85\textwidth]{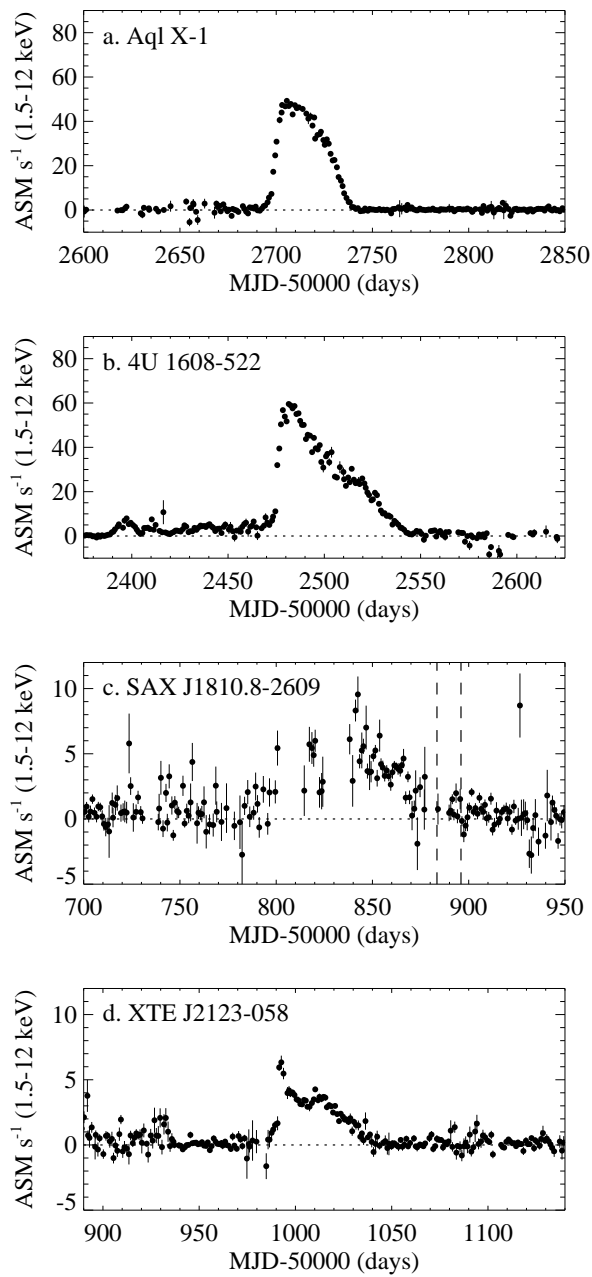}}
\vspace{-13.0cm}
\caption{All-Sky Monitor (ASM) light curves for the four neutron star SXTs in 
our comparison group that had outbursts during the {\em RXTE\/} lifetime.  Each 
point represents the average ASM count rate over 1 day.  One Crab flux equals 
75 ASM counts s$^{-1}$.  For SAX J1810.8--2609 ({\it c}), the vertical {\it dashed 
lines} mark the times of previously reported {\em BeppoSAX\/} and {\em ROSAT} 
observations.  Aql X-1 and 4U 1608--522 have had many outbursts during the lifetime 
of {\em RXTE\/}, and the light curves in ({\it a}) and ({\it b}) are 
representative.\label{fig:asm}}
\end{figure}

Here, we characterize the outburst histories of these five sources to
determine if they are related to the quiescent thermal properties as
is expected in the \cite{bbr98} theory.  From a review of SXT light curves 
covering the time period between 1969 and 1996 \citep{csl97} and the 
{\em RXTE\/} All-Sky Monitor (ASM) covering 1996 to 2003, 4U 1608--522 had 
16 outbursts with a mean peak X-ray luminosity, $\bar{L}_{\rm peak}$, of 
$2.5\times 10^{37}$ ($d$/3.6 kpc)$^{2}$ ergs~s$^{-1}$.  Aql X-1 had 21 
outbursts with $\bar{L}_{\rm peak} = 3.6\times 10^{37}$ ($d$/5 kpc)$^{2}$ 
ergs~s$^{-1}$.  Cen X-4 had two outbursts (in 1969 and 1979) with 
$\bar{L}_{\rm peak} = 5.6\times 10^{37}$ ($d$/1.2 kpc)$^{2}$ ergs~s$^{-1}$, 
and XTE J2123--058 had one outburst (in 1998) with $\bar{L}_{\rm peak} = 
1.7\times 10^{37}$ ($d$/8.5 kpc)$^{2}$ ergs~s$^{-1}$.  Like XTE J2123--058, 
SAX J1810.8--2609 had only one outburst, also in 1998.

SAX J1810.8--2609 was discovered by \cite{natalucci00}; and the only
reported X-ray detections of this source in outburst were from {\em BeppoSAX\/} 
and {\em ROSAT\/} in 1998 March \citep{natalucci00,greiner99}.
However, the ASM light curve shown in Figure~\ref{fig:asm}c indicates
that SAX J1810.8--2609 was in outburst by 1998 January 4 (MJD 50,817)
and probably as early as 1997 December 18 (MJD 50,800).  The dashed lines in 
Figure~\ref{fig:asm}c mark the times of the {\em BeppoSAX\/} discovery at 
MJD 50,882--50,885 and the {\em ROSAT\/} observation at MJD 50,896, while 
it is clear that the source was considerably brighter prior to this.  From 
the 1.5--12 keV ASM light curve, $\bar{L}_{\rm peak} = 1.0\times 10^{37}$ 
($d$/4.9 kpc)$^{2}$ ergs~s$^{-1}$.  Although it has been suggested that 
SAX J1810.8--2609 had an unusually low outburst luminosity \citep{jwv04}, 
the value we derive from the ASM measurements is comparable to the other 
neutron star SXTs.  Figure~\ref{fig:asm} also shows the ASM light curve for 
the XTE J2123--058 outburst as well as sample outburst light curves for
Aql X-1 and 4U 1608--522 (Cen X-4 has not had an outburst during the
{\em RXTE\/} lifetime).  These light curves demonstrate that individual
outbursts from different sources are similar in duration (typically 
40--80 days) and overall shape, although it should be noted that exceptional 
outbursts do occur \citep{csl97}.

For each of the neutron star SXTs, the time-averaged mass accretion 
rate over the past 33~years can be expressed as
\begin{equation}
<\dot{M}> = \frac{\bar{L}_{\rm peak} N t_{\rm outburst} f}{\epsilon c^{2} (33 \rm{yr})} = s \bar{L}_{\rm peak} N~~~~~~~~,
\end{equation}
where $N$ is the number of outbursts, $t_{\rm outburst}$ is the typical 
duration of an outburst from the source, $f$ is a factor with a value 
less than 1.0 that accounts for the shape of the outburst light curve,
$\epsilon$ is the fraction of the rest mass energy released from accreted 
matter, and $s = t_{\rm outburst} f/\epsilon c^{2} (33 \rm{yr})$.  
As the durations and light curve shapes are similar for the sources 
in our comparison group, $s$ is approximately the same for these 
sources, and the time-averaged mass accretion rate over the past 
33~years is proportional to $\bar{L}_{\rm peak}$$N$.  We estimate $s$ 
using $t_{\rm outburst} = 60$~days and $\epsilon = 0.2$, which is the 
value of $\epsilon$ that is typically assumed for accretion onto a neutron
star \citep{rutledge02b}.  Although the precise value of $f$ depends 
on the exact shape of the outburst light curve, it can be approximated 
as the mean outburst flux divided by the peak flux.  For the 
XTE J2123--058 ASM light curve, the mean flux over a 60 day period 
that includes the 1998 outburst is 41\% of the peak level, and we use 
$f = 0.4$, giving $s = 1.1\times 10^{-23}$ s$^{2}$ cm$^{-2}$.  Using
this value of $s$ for all five sources and the values for 
$\bar{L}_{\rm peak}$ and the number of outbursts ($N$) given above, 
we calculate estimates of the time-averaged mass accretion rate
over the past 33~years.  The values for $<$$\dot{M}$$>$ are given in 
Table~\ref{tab:comparison}, and they show that this quantity is a 
relatively good predictor of $L_{\infty}$.  Aql X-1 and 4U 1608--522 
have the the highest values of both $<$$\dot{M}$$>$ and $L_{\infty}$, 
XTE J2123--058 and SAX J1810.8--2609 have the lowest, and Cen X-4 
is intermediate.  The ratio of the value of $<$$\dot{M}$$>$ for Aql X-1 
to that for XTE J2123--058 is 45, which is consistent with their 
ratio of $L_{\infty}$, which is $>38$.  Similarly, the Aql X-1 to 
SAX J1810.8--2609 ratio of $<$$\dot{M}$$>$ is 76, and this is consistent 
with their ratio of $L_{\infty}$, which is $>27$.  While these 
measurements are in line with the \cite{bbr98} theory, a caveat 
is that we cannot be certain that X-ray flux history over the past 
33 years reflects the behavior over the last 10,000 years.

We can use the \cite{bbr98} theory to predict the recurrence time for 
outbursts of XTE J2123--058, assuming that the outbursts are similar.  
Using $L = \epsilon\dot{M} c^{2}$ and the expression relating $L_{\rm q}$ 
to the time-averaged mass accretion rate, $L_{\rm q} = 
9\times 10^{32}$ $<$$\dot{M}$$>_{-11}$~ergs~s$^{-1}$, one obtains 
(see also Wijnands et al. 2001\nocite{wijnands01}) $L_{\rm q} = 
[t_{\rm outburst}/(t_{\rm outburst}+t_{\rm q})]$($<$$L_{\rm outburst}$$>$/130).
Here, $t_{\rm q}$ is the average time the source spends in quiescence
between outbursts, $<$$L_{\rm outburst}$$>$ is the average luminosity 
during an outburst, and $t_{\rm outburst}$ is defined above.  For 
XTE J2123--058, we used the {\em RXTE\/}/ASM light curve shown in 
Figure~\ref{fig:asm}d to determine that the mean 1.5--12 keV count 
rate during the 40 days ($t_{\rm outburst}$) between MJD 50,990 and 
MJD 51,030 was 3.2  s$^{-1}$, corresponding to 
$<$$L_{\rm outburst}$$>$$ = 1.1\times 10^{37}$ ergs~s$^{-1}$ for 
$d = 8.5$ kpc.  From the luminosity upper limit of $L_{\rm q} < 
1.4\times 10^{32}$ ergs~s$^{-1}$, we obtain a lower limit on the
recurrence time, $t_{\rm outburst}+t_{\rm q}$, of 67 years.  The 
known outburst history of XTE J2123--058, one outburst in 33 years, 
assuming that no outbursts were missed, is consistent with these long 
predicted recurrence times.  However, if the recurrence time is in fact 
shorter than 67 years, then a mechanism of enhanced cooling of the core
would be necessary.  The most often mentioned mechanism is neutrino 
cooling of the core due to the direct Urca process, and this mechanism
requires a neutron star with a mass higher than 1.7--1.8\Msun
\citep{colpi01}.  The mass of the neutron star in XTE J2123--058 has 
been measured at $1.5\pm 0.3$\Msun~or 1.04--1.56\Msun, so that the 
mass would have to be at the upper end of the error range for the
direct Urca process to operate.

While we have been focusing primarily on the implications of
the upper limit on the temperature and luminosity of the thermal
component, it is notable that the power-law fit to the spectrum
of XTE J2123--058 also indicates that its X-ray luminosity, at 
$(9^{+4}_{-3})\times 10^{31}$ ergs~s$^{-1}$ in the unabsorbed 0.3--8~keV 
band, is among the lowest for neutron star SXTs.  The quiescent luminosity 
of SAX J1810.8--2609, $\approx 1\times 10^{32}$ ergs~s$^{-1}$ \citep{jwv04}, 
is essentially the same as for XTE J2123--058, and the only other system 
in this class known to have a lower quiescent X-ray luminosity, 
SAX J1808.4--3658, is also the only millisecond X-ray pulsar system 
for which there is a measurement of its quiescent X-ray spectrum. 
The {\em XMM-Newton} spectrum of SAX J1808.4--3658 is well described 
by a power-law with $\Gamma = 1.5$ and a 0.5--10 keV unabsorbed 
luminosity of $5\times 10^{31}$ ergs~s$^{-1}$, and the upper limit
on the contribution from a thermal component is 10\% of the total
X-ray luminosity \citep{campana02}.  If the spectrum of XTE J2123--058 
consists of only a power-law component, then its
$\Gamma = 3.1^{+0.7}_{-0.6}$ is quite a bit softer than that of
SAX J1808.4--3658, and the difference would have to be explained.
On the other hand, if the spectrum for XTE J2123--058 consists of a 
thermal component and a power-law, then its power-law could have the 
same slope as SAX J1808.4--3658.  Concerning the SAX J1808.4--3658 
thermal component, \cite{campana02} conclude that enhanced neutron 
star cooling is required to obtain a quiescent luminosity as low as 
the {\em XMM-Newton} upper limit implies.  However, as mentioned above, 
such conclusions depend on the assumption that the recent outburst 
behavior is indicative of the average mass accretion rate over the 
past 10,000 years.

In summary, our X-ray measurements of XTE J2123--058 in quiescence
show that this system is one of the faintest and least luminous of
the neutron star SXTs.  A comparison of the XTE J2123--058 X-ray 
properties in outburst and quiescence to that of four other neutron 
star SXTs that exhibit short outbursts shows that the systems with 
a higher degree of outburst activity tend to have more luminous 
thermal components in quiescence, and this is in-line with 
predictions of the theory of deep crustal heating.  For 
XTE J2123--058, the upper limit on the thermal luminosity
is consistent with this theory without requiring enhanced neutron 
star cooling if the outburst recurrence time is $>$67 years, which
is consistent with the known outburst history of this source.

\acknowledgments

JAT would especially like to thank Keith Arnaud for his help 
with statistical techniques and XSPEC.  JAT acknowledges useful 
conversations with Tariq Shahbaz and Rudy Wijnands.  JAT and PK
acknowledge partial support from {\em Chandra} awards GO2-3051X 
and GO3-4043X issued by the {\em Chandra} X-ray Observatory Center, 
which is operated by the Smithsonian Astrophysical Observatory 
for and on behalf of NASA under contract NAS8-39073.  DMG 
acknowledges support from a CASS postdoctoral fellowship.

% BIBLIOGRAPHY
%\bibliographystyle{jwapjbib}
%\bibliography{refs}

\clearpage

\begin{thebibliography}{}

\bibitem[\protect\astroncite{{Brown}, {Bildsten} \& {Rutledge}}{1998}]{bbr98}
{Brown}, E.~F., {Bildsten}, L., \& {Rutledge}, R.~E.,  1998, ApJ, 504, L95

\bibitem[\protect\astroncite{{Campana} et~al.}{1998}]{campana98}
{Campana}, S., {Colpi}, M., {Mereghetti}, S., {Stella}, L., \& {Tavani}, M.,
  1998, A\&A~Rev., 8, 279

\bibitem[\protect\astroncite{{Campana} et~al.}{2004}]{campana04}
{Campana}, S., {Israel}, G.~L., {Stella}, L., {Gastaldello}, F., \&
  {Mereghetti}, S.,  2004, ApJ, 601, 474

\bibitem[\protect\astroncite{{Campana} \& {Stella}}{2003}]{cs03}
{Campana}, S., \& {Stella}, L.,  2003, ApJ, 597, 474

\bibitem[\protect\astroncite{{Campana} et~al.}{2002}]{campana02}
{Campana}, S., et~al., 2002, ApJ, 575, L15

\bibitem[\protect\astroncite{{Casares} et~al.}{2002}]{casares02}
{Casares}, J., {Dubus}, G., {Shahbaz}, T., {Zurita}, C., \& {Charles}, P.~A.,
  2002, MNRAS, 329, 29

\bibitem[\protect\astroncite{{Cash}}{1979}]{cash79}
{Cash}, W.,  1979, ApJ, 228, 939

\bibitem[\protect\astroncite{{Chen}, {Shrader} \& {Livio}}{1997}]{csl97}
{Chen}, W., {Shrader}, C.~R., \& {Livio}, M.,  1997, ApJ, 491, 312

\bibitem[\protect\astroncite{{Colpi} et~al.}{2001}]{colpi01}
{Colpi}, M., {Geppert}, U., {Page}, D., \& {Possenti}, A.,  2001, ApJ, 548,
  L175

\bibitem[\protect\astroncite{{Dickey} \& {Lockman}}{1990}]{dl90}
{Dickey}, J.~M., \& {Lockman}, F.~J.,  1990, ARA\&A, 28, 215

\bibitem[\protect\astroncite{{Freeman} et~al.}{2002}]{freeman02}
{Freeman}, P.~E., {Kashyap}, V., {Rosner}, R., \& {Lamb}, D.~Q.,  2002, ApJS,
  138, 185

\bibitem[\protect\astroncite{{Greiner} et~al.}{1999}]{greiner99}
{Greiner}, J., {Castro-Tirado}, A.~J., {Boller}, T., {Duerbeck}, H.~W.,
  {Covino}, S., {Israel}, G.~L., {Linden-V{\o}rnle}, M.~J.~D., \&
  {Otazu-Porter}, X.,  1999, MNRAS, 308, L17

\bibitem[\protect\astroncite{{Haensel} \& {Zdunik}}{1990}]{hz90}
{Haensel}, P., \& {Zdunik}, J.~L.,  1990, A\&A, 227, 431

\bibitem[\protect\astroncite{{Heinke} et~al.}{2003}]{heinke03}
{Heinke}, C.~O., {Grindlay}, J.~E., {Lloyd}, D.~A., \& {Edmonds}, P.~D.,  2003,
  ApJ, 588, 452

\bibitem[\protect\astroncite{{Homan} et~al.}{1999}]{homan99}
{Homan}, J., {M{\' e}ndez}, M., {Wijnands}, R., {van der Klis}, M., \& {van
  Paradijs}, J.,  1999, ApJ, 513, L119

\bibitem[\protect\astroncite{{Hynes} et~al.}{2001}]{hynes01}
{Hynes}, R.~I., {Charles}, P.~A., {Haswell}, C.~A., {Casares}, J., {Zurita},
  C., \& {Serra-Ricart}, M.,  2001, MNRAS, 324, 180

\bibitem[\protect\astroncite{{Jonker}, {Wijnands} \& {van der
  Klis}}{2004}]{jwv04}
{Jonker}, P.~G., {Wijnands}, R., \& {van der Klis}, M.,  2004, MNRAS, 349, 94

\bibitem[\protect\astroncite{{Landolt}}{1992}]{landolt92}
{Landolt}, A.~U.,  1992, AJ, 104, 340

\bibitem[\protect\astroncite{{Levine}, {Swank} \& {Smith}}{1998}]{lss98}
{Levine}, A., {Swank}, J., \& {Smith}, E.,  1998, IAU~Circular, 6955

\bibitem[\protect\astroncite{{Natalucci} et~al.}{2000}]{natalucci00}
{Natalucci}, L., {Bazzano}, A., {Cocchi}, M., {Ubertini}, P., {Heise}, J.,
  {Kuulkers}, E., {in 't Zand}, J.~J.~M., \& {Smith}, M.~J.~S.,  2000, ApJ,
  536, 891

\bibitem[\protect\astroncite{{Rutledge} et~al.}{1999}]{rutledge99}
{Rutledge}, R.~E., {Bildsten}, L., {Brown}, E.~F., {Pavlov}, G.~G., \&
  {Zavlin}, V.~E.,  1999, ApJ, 514, 945

\bibitem[\protect\astroncite{{Rutledge} et~al.}{2001}]{rutledge01}
{Rutledge}, R.~E., {Bildsten}, L., {Brown}, E.~F., {Pavlov}, G.~G., \&
  {Zavlin}, V.~E.,  2001, ApJ, 551, 921

\bibitem[\protect\astroncite{{Rutledge} et~al.}{2002a}]{rutledge02a}
{Rutledge}, R.~E., {Bildsten}, L., {Brown}, E.~F., {Pavlov}, G.~G., \&
  {Zavlin}, V.~E.,  2002a, ApJ, 577, 346

\bibitem[\protect\astroncite{{Rutledge} et~al.}{2002b}]{rutledge02b}
{Rutledge}, R.~E., {Bildsten}, L., {Brown}, E.~F., {Pavlov}, G.~G., {Zavlin},
  V.~E., \& {Ushomirsky}, G.,  2002b, ApJ, 580, 413

\bibitem[\protect\astroncite{{Shahbaz} et~al.}{2003}]{shahbaz03}
{Shahbaz}, T., {Zurita}, C., {Casares}, J., {Dubus}, G., {Charles}, P.~A.,
  {Wagner}, R.~M., \& {Ryan}, E.,  2003, ApJ, 585, 443

\bibitem[\protect\astroncite{{Soria}, {Wu} \& {Galloway}}{1999}]{swg99}
{Soria}, R., {Wu}, K., \& {Galloway}, D.~K.,  1999, MNRAS, 309, 528

\bibitem[\protect\astroncite{{Tavani}}{1991}]{tavani91}
{Tavani}, M.,  1991, ApJ, 379, L69

\bibitem[\protect\astroncite{{Tomsick} et~al.}{1999}]{tomsick99}
{Tomsick}, J.~A., {Halpern}, J.~P., {Kemp}, J., \& {Kaaret}, P.,  1999, ApJ,
  521, 341

\bibitem[\protect\astroncite{{Tomsick} et~al.}{2001}]{tomsick01}
{Tomsick}, J.~A., {Heindl}, W.~A., {Chakrabarty}, D., {Halpern}, J.~P., \&
  {Kaaret}, P.,  2001, ApJ, 559, L123

\bibitem[\protect\astroncite{{Tomsick} et~al.}{2002}]{tomsick02}
{Tomsick}, J.~A., {Heindl}, W.~A., {Chakrabarty}, D., \& {Kaaret}, P.,  2002,
  ApJ, 581, 570

\bibitem[\protect\astroncite{{Wijnands} et~al.}{2003a}]{wijnands03a}
{Wijnands}, R., {Heinke}, C.~O., {Pooley}, D., {Edmonds}, P.~D., {Lewin},
  W.~H.~G., {Grindlay}, J.~E., {Jonker}, P.~G., \& {Miller}, J.~M.,  2003a,
  astro-ph/0310144

\bibitem[\protect\astroncite{{Wijnands} et~al.}{2003b}]{wijnands03b}
{Wijnands}, R., {Homan}, J., {Miller}, J.~M., \& {Lewin}, W.~H.~G.,  2003b,
  astro-ph/0310612

\bibitem[\protect\astroncite{{Wijnands} et~al.}{2001}]{wijnands01}
{Wijnands}, R., {Miller}, J.~M., {Markwardt}, C., {Lewin}, W.~H.~G., \& {van
  der Klis}, M.,  2001, ApJ, 560, L159

\bibitem[\protect\astroncite{{Wilson} et~al.}{2003}]{wilson03}
{Wilson}, C.~A., {Patel}, S.~K., {Kouveliotou}, C., {Jonker}, P.~G., {van der
  Klis}, M., {Lewin}, W.~H.~G., {Belloni}, T., \& {M{\' e}ndez}, M.,  2003,
  ApJ, 596, 1220

\bibitem[\protect\astroncite{{Zavlin}, {Pavlov} \& {Shibanov}}{1996}]{zps96}
{Zavlin}, V.~E., {Pavlov}, G.~G., \& {Shibanov}, Y.~A.,  1996, A\&A, 315, 141

\bibitem[\protect\astroncite{{Zurita} et~al.}{2000}]{zurita00}
{Zurita}, C., et~al., 2000, MNRAS, 316, 137

\end{thebibliography}

% TABLES

\clearpage

\begin{table}
\caption{XTE J2123--058 X-Ray and Optical Observations\label{tab:obs}}
\begin{minipage}{\linewidth}
\footnotesize
\begin{tabular}{lcccccc} \hline \hline
Observatory & UT Date & Energy Band & Exposure Time (s)\\ \hline \hline
{\em BeppoSAX\/} & 2000 May 12--13 & $0.1-10$ keV & 46,340\footnote{This is the MECS exposure 
time.  The LECS exposure time is 17,080~s.}\\
MDM            & 2000 July 24 & Harris $R$ & $3\times 600$\\
Lick           & 2002 September 11 & Kron-Cousins $R$ & $2\times 700$\\
{\em Chandra}  & 2002 November 13 & $0.3-8$ keV & 17,706\\
MDM            & 2002 November 25 & Harris $R$ & $8\times 600$\\
\end{tabular}
\end{minipage}
\end{table}

\begin{table}
\caption{Spectral Fits\label{tab:spectra}}
\begin{minipage}{\linewidth}
\footnotesize
\begin{tabular}{lcccccc} \hline \hline
Model
%\footnote{pl = power-law, bb = blackbody}
& $N_{\rm H}$\footnote{Errors are 90\% confidence for all parameters.} &  $\Gamma$ &
$F_{\rm pl}$\footnote{0.3--8 keV unabsorbed flux.  At $d=8.5$~kpc,
$F_{\rm pl}=10^{-14}$ ergs~cm$^{-2}$~s$^{-1}$ corresponds to $8.6\times 10^{31}$ ergs~s$^{-1}$.} & $kT_{\rm bb}$ & 
$L$/$d_{8.5}^{2}$\footnote{Bolometric luminosity at $d=8.5$~kpc.} & Fit\\
& ($10^{20}$ cm$^{-2}$) & & ($10^{-14}$ ergs~cm$^{-2}$ s$^{-1}$) & 
(eV) & ($10^{32}$ ergs~s$^{-1}$) & Quality\footnote{Fraction of 10,000 simulated spectra for which the C-statistic was
better than the C-statistic obtained for the actual spectrum.}\\ \hline \hline
pl & $7.0^{+40.0}_{-7.0}$ & $3.1^{+2.8}_{-0.8}$ & $1.2^{+31.9}_{-0.5}$ & --- & --- & 0.69\\
bb & $0.0^{+40.0}_{-0.0}$ & --- & --- & $260^{+50}_{-110}$ & $0.49^{+2.76}_{-0.12}$ & 0.97\\
bb+pl & $8.0^{+38.0}_{-8.0}$ & $2.5^{+2.9}_{-2.3}$ & $0.70^{+3.67}_{-0.48}$ & $100^{+130}_{-60}$ & $0.9^{+160}_{-0.7}$ & 0.66\\ \hline
pl & 6.0 & $3.1^{+0.7}_{-0.6}$ & $1.1^{+0.5}_{-0.4}$ & --- & --- & 0.69\\
bb & 6.0 & --- & --- & $240^{+60}_{-50}$ & $0.63^{+0.22}_{-0.20}$ & 0.987\\
bb+pl & 6.0 & $2.5^{+2.1}_{-2.5}$ & $0.7^{+0.5}_{-0.5}$ & $110^{+180}_{-90}$ & $0.6^{+36.9}_{-0.5}$ & 0.64\\
\end{tabular}
\end{minipage}
\end{table}

\begin{table}
\caption{NSA plus Power-law Fits\label{tab:nsa}}
\begin{minipage}{\linewidth}
\footnotesize
\begin{tabular}{cccccc} \hline \hline
$d$ & 
$kT_{\rm eff}$ & 
$F_{\rm NSA}$\footnote{0.3--8 keV unabsorbed flux.} &
$L_{\infty}$\footnote{Bolometric luminosity as seen by a distant observer using 
$4\pi R^{2}\sigma T_{\rm eff}^{4} (1-2GM/Rc^{2})$, where $R$ = 10 km and $M$ = 1.4\Msun.} &
$\Gamma$   & 
$F_{\rm pl}$\footnote{0.3--8 keV unabsorbed flux.  At $d=8.5$~kpc, $F_{\rm pl}=
10^{-14}$ ergs~cm$^{-2}$~s$^{-1}$ corresponds to $8.6\times 10^{31}$ ergs~s$^{-1}$.}\\
(kpc) & (eV) & ($10^{-14}$ ergs~cm$^{-2}$ s$^{-1}$) &($10^{32}$ ergs~s$^{-1}$) &  & ($10^{-14}$ ergs~cm$^{-2}$ s$^{-1}$)\\ \hline\hline
6   &  $< 57$ & $< 1.1$ & $< 0.80$ & $2.3^{+1.4}_{-2.5}$ & $0.56^{+0.82}_{-0.42}$\\
8.5 &  $< 66$ & $< 1.1$ & $< 1.4$ & $2.5^{+1.4}_{-3.1}$ & $0.63^{+0.82}_{-0.60}$\\
11  &  $< 73$ & $< 1.0$ & $< 2.1$ & $2.8^{+1.9}_{-3.4}$ & $0.77^{+0.65}_{-0.77}$\\
\end{tabular}
\end{minipage}
\tablecomments{Neutron Star Atmosphere (NSA) model of Zavlin et al. (1996)\nocite{zps96}, with $N_{\rm H} = 6\times 10^{20}$ cm$^{-2}$
held fixed.}

\end{table}

\begin{table}
\caption{NSA Parameters for Short Outburst Field NS SXTs and X-Ray Activity\label{tab:comparison}}
\begin{minipage}{\linewidth}
\footnotesize
\begin{tabular}{lcccccc} \hline \hline
Source\footnote{Parameters for sources other than XTE J2123--058 are from 
\cite{rutledge99,rutledge01,rutledge02a} and \cite{jwv04}.}       &   $d$  & $kT_{\rm eff}$  &   $R$  & $L_{\infty}$\footnote{Bolometric luminosity as seen by 
a distant observer calculated using $4\pi R^{2}\sigma T_{\rm eff}^{4} g^{2}$.}  &  $g$\footnote{$g = \sqrt{1-2GM/Rc^{2}}$ assuming $M = 1.4\Msun$.} & 
$<$$\dot{M}$$>$\footnote{The time-averaged mass accretion rate over the past 33 years, estimated 
according to $<$$\dot{M}$$> = s \bar{L}_{\rm peak} N$, where $s = 1.1\times 10^{-23}$ 
s$^{2}$ cm$^{-2}$ (see Equation 1).}  \\
 &  (kpc) &   (eV)      &   (km) & (ergs~s$^{-1}$) &  & (\Msun~yr$^{-1}$) \\ \hline \hline
Aql X-1        &  5    &    $94-108$\footnote{The range of values found in three separate {\em Chandra} observations.} &  13.2  & $(5.3-9.4)\times 10^{33}$ & 0.828 & $1.3\times 10^{-10}$\\
4U 1608--522   & 3.6   &  $170\pm 30$ &   9.4   & $5.3\times 10^{33}$ & 0.748 & $6.9\times 10^{-11}$\\
Cen X-4        & 1.2   &  $76\pm 7$   &   12.9  & $4.8\times 10^{32}$ & 0.824 & $1.9\times 10^{-11}$\\
SAX J1810.8--2609 & 4.9 & $< 72$ & 10 & $< 2.0\times 10^{32}$ & 0.765 & $1.7\times 10^{-12}$\\
XTE J2123--058 & 8.5   &  $< 66$ & 10 & $< 1.4\times 10^{32}$ & 0.765 & $2.9\times 10^{-12}$\\
\end{tabular}
\end{minipage}
\end{table}

\end{document}